\documentclass[a4paper,10pt]{article}

\usepackage{pstricks}
\usepackage{graphicx,psfrag}

\usepackage[centertags]{amsmath}
\usepackage{amsfonts}
\usepackage{amssymb}
\usepackage{amsthm}
\usepackage{newlfont}
\usepackage{graphicx}
\usepackage{enumerate}

\newtheorem{theorem}{Theorem}

\newtheorem{lemma}{Lemma}

\theoremstyle{remark}

\theoremstyle{definition}

\theoremstyle{definition}

     \newcommand {\beq}  {\begin{equation}}
      \newcommand {\eeq}  {\end{equation}}

\author{ E.Lakshtanov\thanks{Department of Mathematics, Aveiro University, Aveiro 3810, Portugal.  This work was supported by {\it FEDER} funds through {\it COMPETE}--Operational Programme Factors of Competitiveness (``Programa Operacional Factores de Competitividade'') and by Portuguese funds through the {\it Center for Research and Development in Mathematics and Applications} (University of Aveiro) and the Portuguese Foundation for Science and Technology (``FCT--Fund\c{c}\~{a}o para a Ci\^{e}ncia e a Tecnologia''), within project PEst-C/MAT/UI4106/2011 with COMPETE number FCOMP-01-0124-FEDER-022690, and by the FCT research project
PTDC/MAT/113470/2009 (lakshtanov@rambler.ru)},  B.Vainberg\footnote{Department of
Mathematics and Statistics, University of North Carolina,
Charlotte, NC 28223, USA, brvainbe@uncc.edu   The work was
partially supported  by the NSF grant DMS-1008132.}}

\title{A priori estimates for high frequency scattering by obstacles of arbitrary shape}

\begin{document}
\date{}
\maketitle

\begin{abstract}
High frequency estimates for the Dirichlet-to-Neumann and Neumann-to-Dirichlet operators are obtained for the Helmholtz equation in the exterior of bounded obstacles. These a priori estimates are used to study the scattering of plane waves by an arbitrary bounded obstacle and to prove that the total cross section of the scattered wave does not exceed four geometrical cross sections of the obstacle in the limit as the wave number $k\to \infty$. This bound of the total cross section is sharp.
\end{abstract}

\section{Introduction}

\textbf{High frequency estimate of the Dirichlet-to-Neumann operator.} Let $\Omega$ be the exterior of a bounded obstacle $\mathcal O\subset \mathbb R^3$ with a Lipschitz boundary. Consider the solution $u=u(r)\in H^1_{loc}(\Omega)$ of the Helmholtz equation
\begin{equation}\label{helm}
\Delta u(r)+k^2 u(r)=0, \quad r=(x,y,z) \in \Omega=\mathbb R^3 \backslash
\mathcal O,\quad k > 0,
\end{equation}
in $\Omega$, which satisfies the radiation condition
\begin{equation}\label{Somm}
\int_{|r|=R} \left |\frac{\partial u(r)}{\partial |r|}-iku(r)
\right |^2 dS = o(1), \quad R \rightarrow \infty,
\end{equation}
and the Dirichlet or Neumann boundary condition on $\partial \Omega$
\begin{equation}\label{bcOmega}
\begin{array}{cl}
(D) & u=f\in H^{1/2}(\partial\Omega),\\
(N) & \frac{\partial u}{\partial n} = g \in H^{-1/2}(\partial\Omega).
\end{array}
\end{equation}
Here $n$ is the outer
normal for $\mathcal O$ (it is directed into $\Omega$), which is defined almost everywhere on $\partial\Omega$. The solution $u\in H^1_{loc}(\Omega)$ of problem (\ref{helm})-(\ref{bcOmega}) is understood in the weak sense (it is defined by the Dirichlet form), it exists and is unique (see \cite{McL}). For example, $ u\in H^{1}(\Omega)$ is the solution of the Neumann problem if (\ref{helm}), (\ref{Somm}) hold and
\[
\int_\Omega[-\nabla u\nabla v+k^2uv]dx+\int_{\partial\Omega}gvdS=0
\]
for any $v\in H^{1}(\Omega)$ which vanishes in a neighborhood of infinity. When $\partial\Omega,~f,~v$ are smooth enough, the weak solution belongs to $H^2_{loc}(\Omega).$

In particular, we will consider the scattering of the plane wave $e^{ikr\alpha}, \alpha \in S^2,$ by the obstacle
 $\mathcal O$. Then the scattered wave $u$ satisfies (\ref{helm})-(\ref{bcOmega}) with $f=-e^{ikr\alpha}$ in the case of the Dirichlet problem or $g=-\frac{\partial }{\partial n} e^{ikr\alpha}$ in the case of the Neumann boundary condition.

Every solution $u(r)$  of
(\ref{helm})-(\ref{bcOmega})   has the following behavior at
infinity
\begin{equation}\label{scamp}
u(r)=\frac{e^{ik|r|}}{|r|} u_{\infty}(\theta)+o \left
(\frac{1}{|r|} \right ), \quad r \rightarrow \infty, \quad
\theta=r/|r| \in S^2,
\end{equation}
where the function $u_\infty(\theta)=u_\infty(\theta,k)$ is called
the {\it scattering amplitude} and the quantity
$$
\sigma(k)=\|u_\infty\|^2_{L_2(S^2)}=\int_{S^2}
|u_\infty(\theta)|^2 d\mu (\theta)
$$
is called the {\it total cross section}. Here $d\mu$ is the surface element of
the unit sphere.

Problem (\ref{helm})-(\ref{bcOmega}) can be easily reformulated in terms of the Neumann-to-Dirichlet operator
$\mathcal D= \mathcal D(k)$:
\begin{equation} \label{operD}
\mathcal D(k):H^{-1/2}(\partial \Omega)\rightarrow H^{1/2}(\partial \Omega),~~k \in R,
\end{equation}
which maps the normal derivative $\frac{\partial u}{\partial n }|_{\partial
\Omega}$ of the solution $u\in H^{1}_{loc}(\Omega)$ of the Neumann problem (\ref{helm})-(\ref{bcOmega}) into the value $u |_{\partial\Omega}$ of
the solution at the boundary.
When $k$ is complex, this operator is defined as the meromorphic extension of (\ref{operD}). This extension can be found in \cite{vain68, vain75} in the case of domains with smooth boundaries, but the constructions in \cite{vain68, vain75} remain valid for Lipschitz domains.

Our first result concerns the high frequency estimate of operators $\mathcal D(k)$ and $\mathcal D^{-1}(k)$ in the case of a smooth enough non-trapping obstacle. Recall that an obstacle with a smooth boundary is called non-trapping if an arbitrary geometrical optics ray coming from outside (with the reflection angles equal to the incident angles) goes to infinity.

\begin{theorem}\label{TheorDtNEst}
Let $\mathcal O$ be a non-trapping obstacle with an infinitely (for simplicity) smooth boundary $\partial \Omega$. Then there exists a positive $k$-independent constant $C$ such that
\begin{equation}\label{dinf}
\|\mathcal D \| < Ck , \quad \| \mathcal D^{-1} \| < C k^{3}, \quad k>1.
\end{equation}
\end{theorem}
\textbf{Remark.} The proof of this theorem is based on a reduction to a similar result by one of the authors \cite{vain75} on the resolvent estimates for problem  (\ref{helm})-(\ref{bcOmega}) with an inhomogeneity in the right-hand side of the equation, not in the boundary condition. Note that the estimates in \cite{vain75} are sharp while here we do not care about the sharpness of the estimates (\ref{dinf}).

Let us provide an important consequence of Theorem \ref{TheorDtNEst} which allows one to estimate the accuracy of an approximate solution of the scattering problem and the accuracy of the total cross section when the boundary condition is satisfied approximately. Let $u^D, u^N$ be the scattered fields for the incident plane wave in the case of the Dirichlet or Neumann boundary conditions, respectively, i.e.,  $u^D, u^N$ satisfy  (\ref{helm}), (\ref{Somm}) and
\begin{equation}\label{vvf}
u^{D}=-e^{ik(r \alpha)}, \quad \frac{\partial u^{N}}{\partial n}=-\frac{\partial e^{ik(r \cdot \alpha)}}{\partial n} ,
\quad x\in \partial \Omega.
\end{equation}
Let $u^{D,1},u^{N,1}$ be approximations to the scattered fields which satisfy (\ref{helm}),(\ref{Somm}) and satisfy boundary conditions with some error:
\begin{equation}\label{vvf1}
u^{D,1}=-e^{ik(r \alpha)}+ f, \quad \frac{\partial u^{N,1}}{\partial n}=-\frac{\partial e^{ik(r \cdot \alpha)}}{\partial n} + g.
\end{equation}
Then, for $k>1$,
$$
\begin{array}{ll}
\left \|\frac{\partial u^{D,1}}{\partial n}-\frac{\partial u^{D}}{\partial n} \right  \|_{H^{-1/2}(\partial\Omega)} \leq C k^{3}\|f\|_{H^{1/2}(\partial\Omega)},
\\
\|u^N-u^{N,1}\|_{H^{1/2}(\partial\Omega)} < Ck  \|g\|_{H^{-1/2}(\partial\Omega)},
\end{array}
$$
and
$$
\begin{array}{ll}
\|u^{D,1}_\infty - u^D_\infty \|_{L_2(S^2)} \leq C  k\|f\|_{H^{1/2}(\partial\Omega)}
\\
 \|u^N_\infty-u^{N,1}_\infty\|_{L_2(S^2)} \leq C  \|g\|_{H^{-1/2}(\partial\Omega)}.
\end{array}
$$
The above estimates on the boundary $\partial\Omega$ are the direct consequence of (\ref{dinf}), and the estimates of the scattering amplitudes follow immediately after that from the Green formula.

The next statement holds for arbitrary obstacles which are not necessarily non-trapping or have smooth boundary.
\begin{theorem}\label{TheorTrappedEst}
Let $\mathcal O$ be a bounded obstacle with a Lipschitz boundary.
Then for each $\delta>0$ there exists a positive constant $C=C(\delta)$ such that
\begin{equation}\label{ImWNEst}
\|\mathcal D\| < C\frac{k}{\Im k} , \quad   \| \mathcal D^{-1} \| < C \frac{|k|^{3}}{\Im k},\quad 0< \arg k< \frac{\pi}{2}-\delta, \quad |k|>1.
\end{equation}
\end{theorem}
This result allows one to estimate the error of approximations of the scattered field and the total cross section after averaging them with respect to the wave number $k$. Let $u^{D,1},u^{N,1}$ be the approximations of the scattered fields which satisfy (\ref{helm}),(\ref{Somm}) and (\ref{vvf1}). We assume that the Dirichle/Neumann values of these approximations are analytic in $k$ in a neighborhood of the real $k$-axis as elements of the corresponding functional spaces, and
\begin{equation}\label{bcap}
\|f \|_{H^{1/2}(\partial\Omega)}=O (k^{-m}), \quad\text{or/and} \quad \|g \|_{H^{-1/2}(\partial\Omega)}=O  (k^{-n}),
\end{equation}
when $\Re k\to \infty,\quad |\Im k|\leq c_0<\infty$. The following result is a consequence of Theorem \ref{TheorTrappedEst} (here we only formulate the estimates for the cross sections).
\begin{theorem}\label{cor1}
Let function $u^{D,1}$ or $u^{N,1}$ satisfy (\ref{helm}),(\ref{Somm}), (\ref{vvf1}) and (\ref{bcap}).
Then, for arbitrary positive $\alpha=\alpha(k),~0<\alpha(k)\leq \alpha_0<\infty,$ we have
$$
\int_{k-\alpha(k)}^{k+\alpha(k)} \|u^D_\infty(k')-u^{D,1}_\infty(k')\|^2dk' = O(k^{2-2m}), \quad k \rightarrow \infty,
$$
or/and (in accordance with (\ref{bcap}))
$$
\int_{k-\alpha(k)}^{k+\alpha(k)} \|u^N_\infty(k')-u^{N,1}_\infty(k')\|^2dk' =O(k^{-2n}), \quad k \rightarrow \infty.
$$
\end{theorem}

Finally, let us note that all the high frequency estimates known so far were obtained for the star-shaped obstacles \cite{ML},\cite{babich}, or for the obstacles with a single reflection of the rays \cite{alb} and, later, for smooth non-trapping obstacles and general elliptic equations \cite{vain75}.

{\bf Upper bound for the total cross section when $k \to \infty$.} Consider the scattering of plane waves by an obstacle. Recall that the geometrical cross section $\Theta$ of the obstacle is the shadow of the obstacle illuminated by the plane wave.
In the case of a smooth strictly convex obstacle, it is well known \cite{majdaTay} that the total cross section $\sigma(k)$ at high frequencies coincides with the doubled geometrical cross section, i.e., $\sigma(k) \to 2 \Theta$ as $k\to\infty.$ In fact, $\sigma(k)$ measures the energy of the difference between the unperturbed field (the incident wave) and the field in the presence of the obstacle. One contribution to that difference comes from the fact that the field in the presence of the obstacle is practically zero in the shadow zone (when $k\to\infty$) while the unperturbed field has amplitude one there. The second contribution comes from the wave reflected from the obstacle according to the law of geometrical optics. The arguments of this type can be found in many physics textbooks.

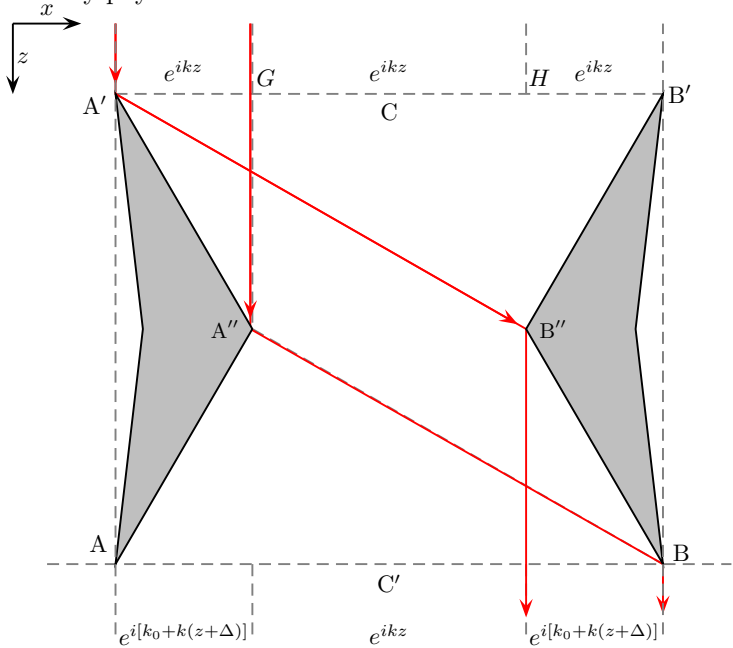
\begin{figure}[h]
\begin{picture}(0,240)
\scalebox{0.9}{ \rput(8.7,5){

\psline[linecolor=gray,linestyle=dashed](-2,4.5)(-2,3.4641)(-2,0)(4,-3.4641)

\psline[linecolor=gray,linestyle=dashed](2,4.5)(2,3.4641)
\psline[linecolor=gray,linestyle=dashed](2,-3.4641)(2,-4.5)
\psline[linecolor=gray,linestyle=dashed](-2,-3.4641)(-2,-4.5)

\rput(3,3.8){$e^{ikz}$}
\rput(0,3.8){$e^{ikz}$}
\rput(-3,3.8){$e^{ikz}$}

\rput(3,-4.5){$e^{i[k_0+k(z+\Delta)]}$} \rput(0,-4.5){$e^{ikz}$}
\rput(-3,-4.5){$e^{i[k_0+k(z+\Delta)]}$}

\psline[linecolor=gray,linestyle=dashed](-5,-3.4641)(5,-3.4641)
\psline[linecolor=gray,linestyle=dashed](-4,3.4641)(4,3.4641)

 \psline[linecolor=red,arrows=->,arrowscale=1.8](-4,4.5)(-4,3.6)
\psline[linecolor=red,arrows=->,arrowscale=1.8](-4,4.5)(-4,3.4641)(1.88,0.069282)
\psline[linecolor=red,arrows=->,arrowscale=1.8](-4,3.4641)(2,0)(2,-4.25)
 \psline[linecolor=red,arrows=->,arrowscale=1.8](-2.03,4.5)(-2.03,0.15)
\psline[linecolor=red,arrows=->,arrowscale=1.8](-2.03,4.5)(-2.03,0)(4,-3.4641)(4,-4.2)

\pspolygon[fillstyle=solid,fillcolor=lightgray](-4,-3.4641)(-3.6,0)(-4,3.4641)(-2,0)
\pspolygon[fillstyle=solid,fillcolor=lightgray](4,-3.4641)(3.6,0)(4,3.4641)(2,0)
\psline[linecolor=gray,linestyle=dashed](-4,-4.5)(-4,4.5)
\psline[linecolor=gray,linestyle=dashed](4,-4.5)(4,4.5)

\rput(-4.25,-3.1641){A} \rput(-4.3,3.3){A$'$}

\rput(4.27,-3.3){B}
\rput(4.25,3.464){B$'$}
\rput(-1.8,3.7){$G$}
\rput(2.2,3.7){$H$}
\rput(-2.4,0){\small A$''$}
\rput(2.4,0){\small B$''$}

   \rput(0,3.2){C}
   \rput(0,-3.75){C$'$}

 \psline[arrows=->,arrowscale=1.8](-5.5,4.5)(-4.5,4.5)
 \rput(-5,4.7){$x$}
 \psline[arrows=->,arrowscale=1.8](-5.5,4.5)(-5.5,3.46)
 \rput(-5.35,4){$z$}
} }
\end{picture}
\label{fig basic construction2} \caption{The Alexenko-Plakhov obstacle $\mathcal O$ is the translation of the two dimensional object above by distance one along the $y$ direction. Here $|AB|=1, |A''B''|=1/2, |AA'|=\sqrt3/2$. The incident plane wave $e^{ikz}$ comes down along the $z$-axis. The geometrical cross section is $1/2.$
The eikonal approximation $\Psi^{eik}$ is equal to $e^{ikz}$ above $A'B'$ and  $e^{ikz}$ or $e^{i[k_0+k(z+\Delta)]}$ below $AB$ with $\Delta=|GA''|+|A''B|-|A'A|$ being a constant. The latter indicates that $\lim_{n \rightarrow \infty} \sigma(k_n)=2(=4\Theta) ~~\text{for}~ k_n=\frac{k_0+(2n +1) \pi}{\Delta},~~\mathbb Z_+\ni n \to\infty
.$}
\end{figure}

Recently we studied \cite{LSV} the scattering of the plane waves by the Alexenko-Plakhov object \cite{alex1} which was suggested as a candidate for an invisible body on the basis of a structure of geometrical optics rays. We proved that the total cross section $\sigma(k)$ for that obstacle approaches four geometrical cross sections $\Theta$ for some sequence $k=k_n\to\infty.$ This non-convex obstacle $\mathcal O$ is the translation of the two dimensional object on Fig. 1 by distance one along the $y$-axis. The incident plane wave $e^{ikz}$ comes down along $z$-axis. Geometrical optics rays are reflected
twice from the boundary of $\mathcal O$ and continue to propagate parallel to each other in the
same way as if the obstacle was absent. We constructed the geometrical optics (eikonal) approximation to the solution of the scattering problem and justified its validity. It is also shown in \cite{LSV} that the presence of the obstacle changes the incident plane wave (for large $k$) only by creating a constant phase shift along the rays in the shadow zone. The phase shift $\delta$ is equal to $k_0+k\Delta$, where $k_0$ depends on the boundary condition and $\Delta$ depends on the geometry of the obstacle. An important effect appears when $k=k_n=\frac{k_0+(2n+1)\pi}{\Delta}$, where $n$ is an integer, $n\to\infty$. Then $\delta=(2n+1)\pi$, i.e., $e^{i\delta}=-1$, and the difference between the incident wave and the eikonal approximation of the field in the shadow zone is equal to the doubled incident wave. The corresponding squares are related by the factor four, and this leads to the fact that $\sigma(k)\to 4\Theta$ when $k=k_n\to\infty$. A rigorous justification of this fact is given in \cite{LSV} (as well as a proof of almost invisibility of the obstacle when $e^{i\delta}=1$, i.e. $k=\frac{k_0+2n\pi}{\Delta},~n\to\infty$).

It was a big surprise for us to find an obstacle with a total cross section being four times larger than the geometrical cross section in the limit of $k=k_n\to\infty$. The next natural question arises immediately: is there an obstacle for which $\limsup_{k\to\infty}\sigma(k)$ is larger than $4\Theta$? One could expect a positive answer based on the fact that the resonances (poles of the analytical continuation of the resolvent in the half plane $\Im k <0$) can approach the real axis at infinity, \cite{i}. In fact, the answer to this question is negative, and this will be justified in the second part of the present paper. If the non-trapping condition is violated, the negative answer is proved here only after certain averaging. Namely, the following theorem will be proved below.

\begin{theorem}\label{applmain}

1) If a bounded obstacle $\mathcal O$ with an infinitely smooth boundary $\partial\Omega$ is non-trapping, then
\begin{equation}\label{aver1}
\limsup_{k \rightarrow \infty} \sigma(k)  \leq 4 \Theta.
\end{equation}

2) For an arbitrary bounded obstacle with a Lipschitz boundary the following relation holds
\begin{equation}\label{aver}
\limsup_{k \rightarrow \infty} \frac{1}{2\alpha(k)} \int_{k-\alpha(k)}^{k+\alpha(k)} \sigma(k')dk'  \leq 4 \Theta,
\end{equation}
where $\alpha(k)$ is an arbitrary positive function such that $k^{-m}<\alpha(k)<\alpha_0$ for some $m,~\alpha_0<\infty$.
\end{theorem}
\textbf{Remarks.} 1) Note that the averaging in the second case can be taken over intervals of fixed length or intervals shrinking at infinity as an inverse power of $k$.

2) A stronger result than (\ref{aver}) will be obtained. It will be shown that, for each $\varepsilon>0$, the scattered field $u$ (which defines the cross section) can be represented as a sum of two fields, $u=u_0+v$, where the total cross section of the field $u_0$ satisfies (\ref{aver}) with $4(\Theta+\varepsilon)$ in the right hand side, and the average of the cross section $\sigma_v(k)$ of the field $v$ decays at infinity faster then any power of  $k$, i.e.,
\[
 \int_{k-\alpha(k)}^{k+\alpha(k)} \sigma_v(k')dk'  =O(k^{-\infty}), \quad k \to\infty.
\]

The proof of Theorem \ref{applmain} is based on a construction of a particular approximation and using a priory estimates discussed above.

\section{Proofs of a priory estimates}

We start this section by recalling some facts on solutions of the
Dirichlet and Neumann problems in non-smooth domains (see
\cite{McL} for more details). Consider an arbitrary exterior
domain $\Omega=R^3\backslash \overline{\mathcal O}$ with a
Lipschitz boundary. Let $\lambda \notin (-\infty, 0],~h\in
L_2(\Omega),$ and let $u\in H^1(\Omega)$ be the solution of the
Dirichlet problem
\begin{equation}\label{diro}
\Delta u -\lambda u=h, \quad x\in \Omega, \quad \gamma u=u|_{\partial\Omega}=f\in H^{1/2}(\partial\Omega),
\end{equation}
or the Neumann problem
\begin{equation}\label{neuo}
\Delta u -\lambda u=h, \quad x\in \Omega, \quad \gamma_nu=\frac {\partial u}{\partial n}|_{\partial\Omega}=g\in H^{-1/2}(\partial\Omega).
\end{equation}

The Dirichlet problem is well defined since the trace operator
$\gamma : H^1(\Omega)\rightarrow H^{1/2}(\partial\Omega)$ is
bounded. The corresponding trace operator $\gamma_n : H^1(\Omega)\rightarrow H^{-1/2}(\partial\Omega)$ for the
normal derivative is not bounded, but $\gamma_n u$
can be defined for the solutions $u\in H^1(\Omega)$ of the equation $\Delta u -\lambda
u=h\in
L_2(\Omega)$. Namely, if $\partial\Omega$ and $u$ are smooth enough, than
the Green formula implies that
\begin{equation}\label{ddn}
\int _\Omega (\nabla u \nabla v+\lambda uv+hv)dx=
- \int _{\partial\Omega} (\gamma_nu)vds \quad \text{for any } v\in H^1(\Omega).
\end{equation}
If $\partial\Omega$ is only Lipschitz and $u\in H^1(\Omega)$
satisfies the equation $\Delta u -\lambda u=h\in
L_2(\Omega)$, then (\ref{ddn})
is used to define $\gamma_nu\in H^{-1/2}(\partial\Omega)$ in
(\ref{neuo}). One can start with a $v\in H^{1/2}(\partial\Omega)$,
construct an arbitrary bounded extension operator
$\eta:H^{1/2}(\partial\Omega)\rightarrow H^{1}(\Omega)$ and
replace $v$ by $\eta v$ in the left hand side of (\ref{ddn}). Then
(\ref{ddn}) defines a bounded functional $\gamma_nu$ on the space
$H^{1/2}(\partial\Omega)$, i.e., $\gamma_nu\in
H^{-1/2}(\partial\Omega)$. It remains to show that  $\gamma_nu$
does not depend on the choice of $\eta$. The proof of the latter
fact and the following lemma can be found in \cite{McL}.
\begin{lemma}\label{l1a}
Let $\lambda \notin (-\infty, 0]$ and $h\in L_2(\Omega)$. Then
problems (\ref{diro}) and (\ref{neuo}) are uniquely solvable in
$H^1(\Omega)$. Moreover, there exists a constant $C=C(\lambda)$
such that
\[
\|u\|_{H^{1}(\Omega)}\leq C(\|h\|_{L_2(\Omega)}+
\|f\|_{H^{1/2}(\partial\Omega)}),~ ||u||_{H^{1}(\Omega)}\leq
C(\|h\|_{L_2(\Omega)}+\|g\|_{H^{-1/2}(\partial\Omega)})
\]
for the solutions of (\ref{diro}) and (\ref{neuo}), respectively, and
\[
\|\gamma_nu\|_{H^{-1/2}(\partial\Omega)}\leq C\|u\|_{H^{1}(\Omega)}
\]
for the solutions of both problems.
\end{lemma}
In fact, Lemma \ref{l1a} is proved in \cite{McL} (Lemma 4.3 and
Theorem 4.10) in the case of bounded domains (for more general
equations), but the condition $\lambda \notin (-\infty, 0]$ allows
one to apply the same arguments to the equation above.
Alternatively, one can easily obtain Lemma \ref{l1a} using the
same statement for bounded domains and a standard technique based
on a partition of the unity.

{\bf Proof of Theorem \ref{TheorDtNEst}.} Let us recall the
resolvent estimate obtained in \cite{vain75}. Assume that the
obstacle $\mathcal O$ satisfies the assumptions of
Theorem~\ref{TheorDtNEst} ($\mathcal O$ is smooth and
non-trapping). We will say that $u\in H^2(\Omega)$ is a solution
of problem (A) if it satisfies the equation
\[
\Delta u+k^2 u=h\in L_{2},\quad x \in \Omega,
\]
radiation condition (\ref{Somm}), and the homogeneous boundary condition (\ref{bcOmega}):
\[
u|_{\partial\Omega}=0\quad \text{or}\quad  u_n|_{\partial\Omega}=0.
\]
Let $h$ have compact support. To be more exact, let $h=0$ for
$|x|>a$. Consider the restriction of $u$ to a bounded region
$\Omega_b=\Omega \bigcap \{|x|<b\}$. Then for each $a,b>0$, there
is a $k$-independent constant $C=C(a,b)$ such that
\begin{equation}\label{apr}
k\|u\|_{L_2(\Omega_b)}+\|u\|_{H^1(\Omega_b)}+k^{-1}\|u\|_{H^2(\Omega_b)}\leq C\|h\|_{L_2(\Omega)}, \quad k\geq 1.
\end{equation}

This estimate is somewhat similar to the estimate for the
solutions of the elliptic equations with a parameter when the
parameter is outside of the spectrum of the problem, for example
for the solutions of the equation $\Delta u-k^2u=h$. In the latter
case, (\ref{apr}) holds with an extra factor $k$ on the left and
with $\Omega $ instead of $\Omega_b$. A weaker result in our case
(the power of $k$ is smaller, the norms are local and the domain
must be non-trapping) is a trade-off for considering the operator on
the continuous spectrum. Note that (\ref{apr}) holds also for the
analytic continuation of the solution in $k$ in a neighborhood of
the real axis which is widening at infinity logarithmically.

Let us prove the first of the estimates in (\ref{dinf}). Assume
first that $g \in H^{1/2}(\partial\Omega)$. Let $v\in H^2(\Omega)$
be the solution of the axillary problem
\[
\Delta v-k^2v=0, \quad x\in \Omega,  \quad u_n|_{\partial\Omega}=g \in H^{1/2}(\partial\Omega).
\]
The Green formula implies that
\begin{equation}\label{apr1}
\int_\Omega(|\nabla v|^2+k^2|v|^2)dx=-\int_{\partial\Omega}v_n\overline{v}dS\leq \|g\|_{H^{-1/2}(\partial\Omega)}\|v\|_{H^{1/2}(\partial\Omega)}.
\end{equation}
From here and the Sobolev imbedding theorem it follows that
\[
\|v\|_{H^{1/2}(\partial\Omega)}\leq C\|v\|_{H^{1}(\Omega)}\leq C[\|g\|_{H^{-1/2}(\partial\Omega)}\|v\|_{H^{1/2}(\partial\Omega)}]^{1/2}, \quad k\geq 1,
\]
and therefore,
\begin{equation}\label{apr2}
\|v\|_{H^{1/2}(\partial\Omega)}\leq C\|g\|_{H^{-1/2}(\partial\Omega)}, \quad k\geq 1.
\end{equation}
From here and (\ref{apr1}) we obtain that
\begin{equation}\label{apr3}
\|\nabla v\|_{L_2(\Omega)}+k\|v\|_{L_2(\Omega)}\leq C\|g|_{H^{-1/2}(\partial\Omega)}, \quad k\geq 1.
\end{equation}

Let us look now for the solution of the problem
(\ref{helm})-(\ref{bcOmega}) in the form $u=\zeta (x)v+w$, where
$\zeta \in C^\infty(\Omega),~\zeta(x)=0$ for $|x|>a,$ and
$\zeta(x)=1$ for $|x|<a-1$ with $a$ so large that the ball
$|x|<a-2$ contains the obstacle $\mathcal O$. Then $w$ is the
solution of the problem (A) with
\[
h=-2\nabla \zeta \nabla v- (\triangle\zeta) v-2k^2\zeta v,
\]
and
\[
\|h\|_{L_2(\Omega)}\leq Ck \|g\|_{H^{-1/2}(\partial\Omega)}, \quad k\geq 1,
\]
due to  (\ref{apr3}).
From here and (\ref{apr}) it follows that
\[
\|w\|_{H^{1}(\Omega_a)}\leq Ck\|g\|_{H^{-1/2}(\partial\Omega)}, \quad k\geq 1,
\]
and therefore
\[
\|w\|_{H^{1/2}(\partial\Omega)}\leq Ck\|g\|_{H^{-1/2}(\partial\Omega)}, \quad k\geq 1.
\]
This and (\ref{apr2}) imply that
\[
\|\mathcal Dg\|_{H^{1/2}(\partial\Omega)}\leq Ck\|g|_{H^{-1/2}(\partial\Omega)}, \quad g\in H^{1/2}(\partial\Omega), \quad k\geq 1.
\]
By taking the closure in the space $H^{-1/2}(\partial\Omega)$ we
arrive at the first estimate in~(\ref{dinf}).

Let us prove the second estimate in (\ref{dinf}). Let $v\in H^1(\Omega)$ be the solution of the axillary problem
\[
\Delta v-v=0, \quad x\in \Omega,  \quad u|_{\partial\Omega}=f \in H^{1/2}(\partial\Omega).
\]
From Lemma \ref{l1a} it follows that
\begin{equation}\label{1b}
c\|\gamma_nv\|_{H^{-1/2}(\partial\Omega)}\leq \|v\|_{H^1(\Omega)}\leq C\|f\|_{H^{1/2}(\partial\Omega)}.
\end{equation}
We look for the solution of the problem
(\ref{helm})-(\ref{bcOmega}) in the form $u=\zeta (x)v+w$, where
$\zeta $ and $a$ are the same as above. Then $w$ is the solution
of the problem (A) with
\[
h=-2\nabla \zeta \nabla v- (\triangle\zeta) v+(1-k^2)\zeta v.
\]
From (\ref{apr}) and the second inequality in (\ref{1b}) it follows that
\[
\|w\|_{H^{2}(\Omega_a)}\leq Ck^3\|f\|_{H^{1/2}(\partial\Omega)}, \quad k\geq 1.
\]
Then the Sobolev imbedding theorem implies that
\[
\|\gamma_nw\|_{H^{1/2}(\partial\Omega)}\leq Ck^3\|f\|_{H^{1/2}(\partial\Omega)}, \quad k\geq 1.
\]
This and (\ref{1b}) justify the second estimate in (\ref{dinf}).

{\bf Proof of Theorem \ref{TheorTrappedEst}.}
Let $\Im (k^2)> 0$. Let $u$ satisfy (\ref{helm}),(\ref{Somm}) and the Neumann boundary condition (\ref{bcOmega}).
The Green formula
\[
\int_\Omega(-\nabla u\nabla v+k^2uv)dx=\int_{\partial\Omega}(\gamma_nu)vdS
\]
remains valid (see \cite{McL}) for arbitrary $u,v \in H^1(\Omega)$ in a Lipschitz domain $\Omega$ if $\Delta u+k^2u=0$ in $\Omega$. This formula with $v=\overline{u}$ implies that
\begin{equation}\label{2}
\Im (k^2) \int_{\Omega} |u|^2 dx \leq \|\gamma_nu\|_{H^{-1/2}(\partial\Omega)}\|u\|_{H^{1/2}(\partial\Omega)},
\end{equation}
and
\begin{equation*}
\|u\|^2_{H^{1}(\Omega)} \leq \frac {|k|^2}{\Im (k^2)}\|\gamma_nu\|_{H^{-1/2}(\partial\Omega)}\|u\|_{H^{1/2}(\partial\Omega)}, \quad \frac {|k|^2}{\Im (k^2)}>1.
\end{equation*}
Together with the Sobolev imbedding theorem
\[
\|u\|_{H^{1/2}(\partial\Omega)}\leq C\|u\|_{H^{1}(\Omega)},
\]
this leads to
\[
\|u\|_{H^{1/2}(\partial\Omega)}\leq C\frac {|k|^2}{\Im (k^2)}\|\gamma_nu\|_{H^{-1/2}(\partial\Omega)},\quad \frac {|k|^2}{\Im (k^2)}>1.
\]
This estimate proves the first inequality in (\ref{ImWNEst}).

Let us prove the second inequality in (\ref{ImWNEst}).  Let $u$
satisfy (\ref{helm}),(\ref{Somm})
and the Dirichlet boundary
condition (\ref{bcOmega}). Estimate (\ref{2}) is still valid in
this case. We rewrite the equation $\Delta  u+k^2u=0$ in the form
\[
\Delta  u-u=h, \quad h=-(1+k^2)u.
\]
Then from Lemma \ref{l1a} and  (\ref{2}) it follows that
\[
\|\gamma_nu\|_{H^{-1/2}(\partial\Omega)}\leq C (\|h\|_{L_2(\Omega)}+\|u\|_{H^{1/2}(\partial\Omega)})
\]
\[
\leq C\kappa\|\gamma_nu\|^{1/2}_{H^{-1/2}(\partial\Omega)}\|u\|^{1/2}_{H^{1/2}(\partial\Omega)}
+C\|u\|_{H^{1/2}(\partial\Omega)}, \quad \kappa=\frac {1+|k|^2}{\sqrt{\Im (k^2)}}.
\]
We combine the last inequality with
\[
C\kappa\|\gamma_nu\|^{1/2}_{H^{-1/2}(\partial\Omega)}\|u\|^{1/2}_{H^{1/2}(\partial\Omega)}\leq \frac{1}{2}\|\gamma_nu\|_{H^{-1/2}(\partial\Omega)}+\frac{1}{2}C^2\kappa^2\|u\|_{H^{1/2}(\partial\Omega)}
\]
and arrive at
\[
\|\gamma_nu\|_{H^{-1/2}(\partial\Omega)}\leq C_1 \frac {|k|^4}{\Im (k^2)}\|u\|_{H^{1/2}(\partial\Omega)}, \quad |k| >1.
\]
This implies the second inequality in (\ref{ImWNEst}).

The proof of Theorem \ref{TheorTrappedEst} is complete.

\textbf{Proof of Theorem \ref{cor1}.} Let us assume that the
Dirichlet boundary
condition is imposed on the boundary of the
domain; the Neumann condition is treated absolutely similarly.
Denote $v=u^{D}-u^{D,1}$. The Green formula implies
 \begin{equation}\label{112}
 \|v_\infty \|^2= \frac{1}{k} \Im \int_{\partial \mathcal O} \frac{\partial v}{\partial n} \overline{v} dS= \frac{1}{k} \Im \int_{\partial \mathcal O}( {\mathcal D^{-1}}v )\overline{v} dS, \quad k>0.
 \end{equation}

Consider the quadratic polynomial $\phi(z)=(2\alpha_0)^2-z^2,$
where $\alpha_0$ was introduced in the statement of Theorem \ref{cor1}. Then $0<\phi(z)\leq 4\alpha_0$ for
$-2\alpha_0<z<2\alpha_0$. Thus $$ \int_{k-\alpha(k)}^{k+\alpha(k)}
\|v_\infty(k')\|^2dk'\leq C\int_{k-\alpha(k)}^{k+\alpha(k)}
\phi(k'-k)\|v_\infty(k')\|^2dk' $$
\[
<C\int_{k-2\alpha_0}^{k+2\alpha_0} \phi(k'-k)\|v_\infty(k')\|^2dk'
\]
\begin{equation}\label{25}
= \frac{C}{k} \Im \int_{k-2\alpha_0}^{k+2\alpha_0} \phi(k'-k)\int_{\partial \mathcal O}(\mathcal D^{-1}v )\overline{v} dk'dS, \quad k>0.
 \end{equation}
Here $v=v(k')$, considered as a function with values in
$H^{1/2}(\partial\Omega)$, is analytic in the half strip indicated
in (\ref{bcap}). The function $\overline{v(\overline{k'})}$ is an
analytic extension of $\overline{v}(k')$ from the real axis to the
same half strip. The operator $\mathcal
D^{-1}:~H^{1/2}(\partial\Omega)\to H^{-1/2}(\partial\Omega)$ is
analytic in the upper half plane. Thus the segment of integration
$[k-2\alpha_0, k+2\alpha_0]$ in (\ref{25}) can be replaced by the
contour $\Gamma= \Gamma_1 \cup \Gamma_2 \cup \Gamma_3$ (see Fig.
2), where $$ \Gamma_1 =\{ k' \in \mathbb C \, : \, 0\leq \Im k'
\leq c_0,~ \Re k' = k-2\alpha_0 \}, \quad $$ $$ \Gamma_2 =\{ k'
\in \mathbb C \, : \,  \Im k' = c_0, ~k-2\alpha_0 \leq \Re k' \leq
k+2\alpha_0 \}, $$ $$ \Gamma_3 =\{ k'\in \mathbb C \, : \, c_0
\geq \Im k' \geq 0 , ~\Re k' = k+2\alpha_0 \}, $$ with $c_0$
defined in (\ref{bcap}).
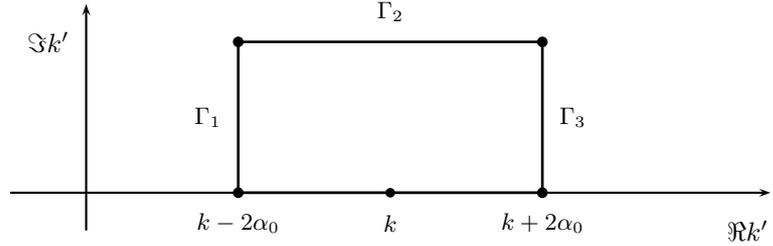
\begin{figure}[h]
\begin{picture}(1.5,80)
\scalebox{1}{ \rput(8,0.5){

\psdot[dotstyle=*](0,0)

 \rput(0,-0.4){\small$k$}
 \rput(2,-0.4){\small$k+2\alpha_0$}
 \rput(-2,-0.4){\small$k-2\alpha_0$}

 \rput(-4.5,2){$\Im k'$}

 \rput(0,2.4){\small$\Gamma_2$}
 \rput(2.4,1){\small$\Gamma_3$}
 \rput(-2.4,1){\small$\Gamma_1$}

\psline{->}(-5,0)(5,0) \pspolygon[linewidth=1pt,
showpoints=true](-2,0)(-2,2)(2,2)(2,0)

\psline{->}(-4,-0.5)(-4,2.5)

 \rput(4.7,-0.5){$\Re k'$}
} }
\end{picture}
\label{CCC} \caption{Contour $\Gamma$}
\end{figure}
Hence,
\[
\int_{k-\alpha(k)}^{k+\alpha(k)} \|v_\infty(k')\|^2dk'\leq \frac{C}{k} \Im\int_{\Gamma} \phi(k'-k) \int_{\partial \mathcal O}(\mathcal D^{-1}v )\overline{v} dk'dS, \quad k>0,
\]
and
\[
\int_{k-\alpha(k)}^{k+\alpha(k)} \|v_\infty(k')\|^2dk'\leq \frac{C}{k}\int_{\Gamma} |\phi(k'-k)|\|(\mathcal D^{-1}v )\|_{H^{-1/2}(\partial\Omega)}\|\overline{v}\|_{H^{1/2}(\partial\Omega)}d|k'|.
\]
It remains to note that contour $\Gamma$ has length $4\alpha_0+2c_0, ~ |\phi(k'-k)|\|\mathcal D^{-1}\|\leq k^3$ on $\Gamma$ due to Theorem \ref{TheorTrappedEst}, and (\ref{bcap}) holds.

The proof of Theorem \ref{cor1} is complete.

\section{The bound for the scattering cross section.}

Let us give an outline of the proof of the bound followed by
rigorous arguments. The proof is based on the introduction of a
specific solution  $\Phi(r)$ of (\ref{helm}) which is a sum of an
incoming and outgoing spherical waves with the amplitudes
$\Phi_\infty^{in}(\theta)$ and $\Phi_\infty^{out}(\theta)$,
respectively, and which has the following properties. Let us put
the obstacle inside of a cylinder $C$ with the axis parallel to
the $z$ axis (which is chosen to be the direction of the incident
wave.) As $k \to \infty,$ function $\Phi(r)$ almost coincides with
the incident plane wave $e^{ikz}$, and it vanishes (as $k \to
\infty$) outside of $C$.  The function  $\Phi(r)$ can be chosen in
such a way that  the total cross section of each component of
$\Phi(r)$ is as close to the geometrical cross section $\Theta$ of the obstacle  $\mathcal O$ as we
please. Due to the a priory estimates obtained in the first part
of the paper, the scattered field $u$ is close to the outgoing
solution $u_0$ of (\ref{helm}) which satisfies the boundary
condition determined by $-\Phi(r)$ (instead of $-e^{ikz}$). Since
$\Phi(r)+u_0$ has zero boundary condition, from the unitarity of
the scattering matrix it follows that
\[
\|\Phi_\infty^{in}\|_{L_2(S^2)}=\|\Phi_\infty^{out}+u_0\|_{L_2(S^2)}.
\]
This implies the bound for the total cross section for $u_0$, and
therefore for $u$.

The rigorous proof starts with the construction of the function
$\Phi$. Denote the $(x,y)$-coordinate plane in $\mathbb R^3$ by
$P.$ Let $\alpha=(0,0,1)$ be the unit normal to $P$, and let $M$
be a domain in $P$ bounded by a polygon $\partial M$. Denote by
$C(M)$ the infinite cylinder with the axis parallel to the
$z$-axis and the cross section $M$, i.e., $$ C(M) = \{ r=r_0 + t
\alpha \in \mathbb R^3 ~  :  \quad r_0 \in M, \quad
-\infty<t<\infty \}, $$

Consider the following function
\begin{equation}\label{DDef}
p(r)=\int_{M} \eta(q) \frac{\sin(k|r-q|)}{|r-q|}
dS(q), \quad r \in \mathbb R^3,
\end{equation}
where $\eta$ is a $C^{\infty}$-function on $P$ with support in
$M$ such that $\eta$ vanishes in a $\delta$-neighborhood of the boundary
$\partial M$ and equals one outside of a $2\delta$-neighborhood of
$\partial M$.
\begin{lemma}\label{lp}
The function $p(r)$ has the following properties

1) $p(r)\in C^\infty (\mathbb R^3)$ and $(\Delta+k^2)p(r)=0$,

2) $p$ is an entire function of $k$,

3) If $K$ is an arbitrary compact on $C(M)$ with the distance from
$\partial C(M)$ at least $2\delta$, then $$ p(r)=2\pi k^{-1}\cos
(kz)+O(k^{-\infty}), \quad r \in K,~~\Re k\to \infty,\quad |\Im
k|\leq c_0<\infty, $$

4)
$$
p(r)=p_\infty^{out}(\theta)\frac {e^{ik|r|}}{|r|}+p_\infty^{in}(\theta)\frac {e^{-ik|r|}}{|r|}+O(|r|^{-2}) \quad \text{as}\quad r \to \infty,
$$
where
\begin{equation}\label{c0}
p_\infty^{out}(\theta)=\frac {1}{2i}\int_M\eta e^{-ik\theta\cdot q}dS(q), \quad p_\infty^{in}(\theta)=\frac {-1}{2i}\int_M\eta e^{ik\theta\cdot q}dS(q).
\end{equation}

5) All the relations above admit differentiation in $r$ of any order.
\end{lemma}
\textbf{Proof.} All the statements above can be derived from the
properties of a single layer obtained in \cite{LSV}. However, this
particular function $p(r)$ is so simple, that it is easer to prove
the lemma independently. Indeed, the first two statements are
obvious since the integrand in (\ref{DDef}) is infinitely smooth,
satisfies the Helmholtz equation and is an entire function of $k$.

In order to prove the third statement, we use the polar
coordinates $\sigma, \phi$ on $P$ with the origin at the point
$r_0=(x,y)$. Then $p(r)$ takes the form
\[
p(r)=\int_0^{2\pi}\int_0^d\eta (r_0+\sigma
\beta)\frac{\sin(k\sqrt{z^2+\sigma^2})}{\sqrt{z^2+\sigma^2}}\sigma
d\sigma d\phi, \quad \beta=(\cos \phi, \sin \phi),
\]
where $d$ is the diameter of $M$. After substitution
$\sqrt{z^2+\sigma^2}=\tau$ and integration by parts, we obtain
\[
p(r)=\int_0^{2\pi}\int_{|z|}^d\eta (r_0+\sqrt{\tau^2-z^2}\beta)\sin(k\tau)d\tau d\phi.
\]
\[
=2\pi k^{-1}\cos(kz)+\int_0^{2\pi}\int_{|z|}^d\frac {d}{d\tau}\eta (r_0+\sqrt{\tau^2-z^2}\beta)k^{-1}\cos(k\tau)d\tau d\phi.
\]

Since $\eta=1$ in a neighborhood of the point $\tau=|z|$, and all
the derivatives of $\eta$ are zeroes at $\tau=0$, further
integration by parts leads to the estimate $O(k^{-\infty})$ for
the last term above. This completes the proof of the third
statement.

To prove the fourth statement, we replace the sine-function in (\ref{DDef}) by
the difference of exponents:
 \[
 \sin k(|r-q|)=\frac{e^{ik|r-q|}-e^{ik|r-q|}}{2i},
 \]
and then use the expansion that is standard in the scattering
theory
 \[
 \frac{e^{ik|r-q|}}{|r-q|}= e^{-ik\theta\cdot q}\frac{e^{ik|r|}}{|r|}+O(|r|^{-2}),\quad r\to\infty, \quad |q|<q_0<\infty.
 \]
The last statement of the lemma is also obvious.

Lemma \ref{lp} implies
\begin{lemma}\label{lp1} The function $\Phi(r)=\frac{k}{2\pi}p(r)-\frac{i}{2\pi}p_z(r)$ has the
following properties

1) $\Phi(r)\in C^\infty (\mathbb R^3)$ and $(\Delta+k^2)\Phi(r)=0$,

2) $\Phi$ is an entire function of $k$,

3) If $K$ is an arbitrary compact on $C(M)$ with the distance from
$\partial C(M)$ at least $2\delta$, then
\begin{equation}\label{ef}
\Phi(r)=e^{ikz}+O(|k|^{-\infty}), \quad r\in K,~~\Re k\to \infty,\quad |\Im k|\leq c_0<\infty,
\end{equation}

4)
$$
\Phi(r)=\Phi_\infty^{out}(\theta)\frac {e^{ik|r|}}{|r|}+\Phi_\infty^{in}(\theta)\frac {e^{-ik|r|}}{|r|}+O(|r|^{-2}) \quad \text{as}\quad r \to \infty,
$$
where
\begin{equation}\label{c0}
\Phi_\infty^{out}(\theta)=k\frac {-1+\theta_3}{4\pi i}\int_M\eta e^{-ik\theta\cdot q}dS(q), \quad \Phi_\infty^{in}(\theta)=-k\frac {1+\theta_3}{4\pi i}\int_M\eta e^{ik\theta\cdot q}dS(q)
\end{equation}
with $\theta_3=z/|r|$.

5) All the relations above admit differentiations in $r$ of any order.
\end{lemma}

\begin{lemma}\label{last}
The following relation holds
\begin{equation}\label{amplit}
\int_{S^2}|\Phi_\infty^{out}(\theta)|^2dS=\int_{S^2}|\Phi_\infty^{in}(\theta)|^2dS=\int_M\eta^2 dS(q)+O(k^{-1}),~ k\to\infty.
\end{equation}
\end{lemma}
\textbf{Proof.} From (\ref{c0}) it follows that $\Phi_\infty^{out}(\theta')=\Phi_\infty^{in}(\theta)$, where $\theta'=(\theta_1,\theta_2,-\theta_3)$. Thus, it is enough to evaluate the first integral above.

For a vector $r=(x,y,z)$, denote by $r'=(x,y)$ its projection on
the $(x,y)$-plane. Obviously,
$|\Phi_\infty^{out}(\theta)|=O(k^{-\infty})$ if $|\theta'|$ is
separated from zero, i.e.,
\[
\Phi_\infty^{out}(\theta)=k\beta(\theta')\frac {\theta_3-1}{2\pi}\int_M\eta e^{-ik\theta\cdot q}dS(q)+O(k^{-\infty})
\]
where $\beta\in C^\infty,~\beta=1$ when $|\theta'|<1/3,$ $\beta=0$ when $|\theta'|>2/3.$ Then with the accuracy of $O(k^{-\infty})$, we have
\[
\int_{S^2}|\Phi_\infty^{out}(\theta)|^2dS=\frac {k^2}{(4\pi)^2}\int_{S^2}\beta(\theta_3-1)^2\int_M\int_M\eta(p')\eta(q') e^{-ik\theta'\cdot (q'-p')}dq'dp'dS
\]
\[
=\frac {k^2}{(2\pi)^2}\int_{|\theta'|<1}\gamma(\theta')\int_M\int_M\eta(p')\eta(q') e^{-ik\theta'\cdot (q'-p')}dq'dp'd\theta',
\]
where
\[
\gamma(\theta')=\frac{\beta(\theta')}{\sqrt{1-|\theta'|^2}}[(\sqrt{1-|\theta'|^2}-1)^2+(\sqrt{1-|\theta'|^2}+1)^2].
\]
The two terms in the last factor above correspond to integration
over the upper and lower hemispheres. We apply the stationary
phase method to the integral with respect to $\theta',q'$. The
only stationary phase point is the point $\theta'=0,q'=p'$, and
this immediately leads to the statement of the lemma. Lemma
\ref{last} is proved.

\textbf{Proof of Theorem \ref{applmain}}. We fix an arbitrary
$\varepsilon >0$ and then choose $M$ in such a way that $M$
contains the projection of the obstacle $\mathcal O$ on the plane
$P$, and the area $|M|$ of $M$ does not exceed
$\Theta+\varepsilon$. Then we choose $\delta$ in the definition of
the function $\eta$ to be so small that $\eta=1$ in a neighborhood
of the projection of the obstacle. Then (\ref{c0}) implies that
\begin{equation}\label{c01}
\Theta\leq\int_M \eta^2dS(q)\leq\Theta+\varepsilon.
\end{equation}

Let $u^0$ satisfy (\ref{helm}),(\ref{Somm})  and one of the boundary conditions
\begin{equation}\label{u0f}
\left (  u^0 +\Phi(r)\right ) |_{\partial \mathcal O} =0 \quad
\text{or} \quad \frac{\partial (u^0  +\Phi(r))}{\partial n}
|_{\partial \mathcal O} =0.
\end{equation}
The choice depends on the boundary condition in (\ref{bcOmega}).
Note that $\Phi$ is a sum of an outgoing and an incoming spherical
waves, and $u^0$ is an outgoing wave. Thus, the Green formula for
$w=u^0_\infty-\Phi^{out}_\infty$ and $\overline{w}$ in the domain
$\Omega_R=\Omega \bigcap \{r:|r|<R\}$ followed by the limiting
process as $R \to\infty$ immediately implies that $$
\|\Phi^{in}_\infty\|=\|u^0_\infty-\Phi^{out}_\infty\|, $$ and
therefore that
\[
\|u^0_\infty\|\leq \|\Phi^{out}_\infty\|+\|\Phi^{in}_\infty\|.
\]
This, Lemma \ref{last} and (\ref{c01}) imply
\begin{equation}\label{vb}
 \|u^0_\infty\|^2\leq 4(\Theta+\varepsilon).
\end{equation}

The latter estimate is very close to the one which is needed to
prove the theorem. Indeed, recall that $u$ is the scattered wave
defined by the incident plane wave, i.e, $u$ satisfies
(\ref{helm}), (\ref{Somm}) and one of the boundary conditions
in~(\ref{vvf}). The function $u^0$ also satisfies (\ref{helm}),
(\ref{Somm}) and the corresponding boundary condition
in~(\ref{u0f}). Then the difference $v=u-u_0$ satisfies
(\ref{helm}), (\ref{Somm}), and from (\ref{ef}) it follows that
\begin{equation}\label{vb}
 \|v\|_{H^{1/2}(\partial \mathcal O)} =O(|k|^{-\infty}) \quad \text{or} \quad \|\frac{\partial v}{\partial n} \|_{H^{-1/2}(\partial \mathcal O)} =O(|k|^{-\infty}),
\end{equation}
when $\Re k\to \infty,\quad |\Im k|\leq c_0<\infty$. The choice
depends on the boundary condition in (\ref{bcOmega}), and,
respectively, in (\ref{vvf}). Since $\varepsilon$ is arbitrary and
$u=u_0+v$, it remains to show that $\|v_\infty\|^2$ vanishes as
$k\to\infty$ if the obstacle is non-trapping and smooth, and
\begin{equation}\label{vest}
\frac{1}{2\alpha(k)}\int_{k-\alpha(k)}^{k+\alpha(k)} \|v_\infty(k')\|^2dk'\to 0, \quad k\to\infty,
\end{equation}
for an arbitrary obstacle with a Lipschitz boundary.

The proofs of both results start with the relation
$$
\|v_\infty \|^2 = \frac{1}{k} \Im \int_{\partial \mathcal O} \frac{\partial v}{\partial n} \overline{v} dS, \quad k>0,
$$
which is an immediate consequence of the Green formula. Let us
assume that the Dirichlet boundary condition is imposed on the
boundary of the domain. The Neumann condition is treated
absolutely similarly. Then
\begin{equation}\label{112a}
\|v_\infty \|^2 = \frac{1}{k} \Im \int_{\partial \mathcal O}( {\mathcal D^{-1}}v )\overline{v} dS, \quad k>0.
\end{equation}

If the obstacle is smooth and non-trapping, then the relation
$\|v_\infty \|=O(k^{-\infty}),$ $ k\to\infty$, follows immediately
from (\ref{112a}), the first estimate in (\ref{vb}), and Theorem
\ref{TheorDtNEst}. Let $\mathcal O$ be an arbitrary obstacle with
a Lipschitz boundary. Then (\ref{vb}) and Theorem \ref{cor1} imply
that
\[
\int_{k-\alpha(k)}^{k+\alpha(k)} \|v_\infty(k')\|^2dk'=O(k^{-\infty}), \quad k\to\infty.
\]

The proof is complete.

\end{document}